\documentclass[%twocolumn,
nofootinbib,
nobibnotes,
amsmath,amssymb,
aps,
prd,
floatfix,
]{revtex4-2}

\usepackage{placeins}
\usepackage{booktabs}
\usepackage{dcolumn}
\usepackage{array}
\usepackage{longtable}
\usepackage{float}
\usepackage{hyphenat}
\usepackage{natbib}
\usepackage{hyperref}

\usepackage{amsmath,amsfonts,amsthm,amssymb}
\usepackage{bm,graphicx,graphics,color}
\usepackage{epsf,epsfig}
\usepackage[all]{xy}
\newcommand*{\B}[1]{\ifmmode\bm{#1}\else\textbf{#1}\fi}

\def\bx{\mathbf{x}}

\def\bp{\mathbf{p}}

\def\no{\nonumber}
\def\lb{\label}
\def\be{\begin{equation}}
\def\ee#1{\label{#1}\end{equation}}
\newcommand{\ben}{\begin{eqnarray}}
\newcommand{\een}{\end{eqnarray}}

\begin{document}

\title{Relativistic Corrections to the Thermodynamic Properties of a Bose-Einstein Gas}

\author{Gilberto M. Kremer}
\email{kremer@fisica.ufpr.br}
\affiliation{Departamento de F\'{i}sica, Universidade Federal do Paran\'{a}, Curitiba 81531-980, Brazil}

%%%%%%%%%%%%%%%%%%%%%%%%%%%%%%%%%%%%%%%%%%%%%%%%%%%%%%%%%%%%%%%
\begin{abstract}
In this work, relativistic corrections to the thermodynamic properties of a Bose-Einstein gas are investigated within the framework of relativistic kinetic theory. The equilibrium Bose-Einstein distribution function is expanded up to order $(1/c^4)$, and the corresponding particle number density, internal energy, and pressure are obtained from the particle four-flow and energy-momentum tensor. Particular attention is given to the first relativistic corrections of order $(kT/mc^2)$. The heat capacity at constant volume is analyzed in the vicinity of the Bose-Einstein condensation temperature, and its behavior below, at, and above the critical temperature is determined. The results show that the specific heat remains continuous at the critical temperature in the non-ultrarelativistic regime. Numerical estimates for Bose-Einstein condensate stars indicate that relativistic effects can increase the specific heat by approximately 5\% at the critical temperature and 4.5\% in the high-temperature limit. Furthermore, a virial expansion is developed to obtain relativistic corrections to the equation of state, specific internal energy, specific heat, chemical potential, enthalpy, and entropy. These results provide a systematic description of relativistic effects in the thermodynamic behavior of Bose-Einstein gases and may be relevant to finite-temperature Bose-Einstein condensate stars.
\end{abstract}

%Relativistic corrections to the thermodynamic properties of a quantum Bose gas are investigated within the framework of the relativistic Boltzmann equation. Starting from the relativistic Bose–Einstein equilibrium distribution, an expansion in powers of the inverse speed of light is performed up to the first relativistic correction, characterized by the parameter $(kT/mc^2)$. The particle number density, internal energy, and pressure are derived from the particle four-flow and energy-momentum tensor. The resulting expressions are used to analyze the specific heat at constant volume in the vicinity of the Bose–Einstein condensation temperature. It is shown that the specific heat remains continuous at the critical temperature, and the corresponding relativistic correction is obtained. Furthermore, a virial expansion is developed in terms of the thermal de Broglie wavelength and specific volume, yielding relativistic corrections to the equation of state, internal energy, specific heat, chemical potential, enthalpy, and entropy. These results provide a systematic description of the leading relativistic effects on the equilibrium thermodynamics of a Bose–Einstein gas.
%\end{abstract}
%
\maketitle 
%%%%%%%%%%%%%%%%%%%%%%%%%%%%%%%%%%%%%%%%%%%%%%%%%%%%

\section{Introduction}
\lb{s.1}

The study of quantum gases has attracted considerable attention due to the fundamental role played by quantum statistics in determining the thermodynamic properties of matter. Among the different quantum statistical systems, the Bose-Einstein gas is of particular interest because it exhibits Bose-Einstein condensation, a phenomenon in which a macroscopic number of particles occupies the lowest energy state when the temperature is below a critical value. This phenomenon has important applications in quantum statistical mechanics and has also been investigated in astrophysical systems, such as Bose-Einstein condensate stars.

The equilibrium distribution function of a relativistic quantum gas was first obtained by J\"uttner \cite{JT} and can be used to describe systems in which the effects of special relativity become relevant. In the framework of relativistic kinetic theory, the thermodynamic properties of a gas can be determined from the one-particle distribution function through the particle four-flow and the energy-momentum tensor \cite{CK}. In the non-relativistic limit, the equilibrium distribution reduces to the familiar quantum statistical distributions, while relativistic effects introduce corrections to the particle number density, pressure, and internal energy.

Relativistic corrections to the thermodynamic properties of ideal gases have been extensively studied for Maxwell-Boltzmann and Fermi-Dirac statistics (see e.g. \cite{CK}). However, the corresponding corrections for Bose-Einstein statistics are particularly relevant near the condensation temperature, where quantum statistical effects become significant. The behavior of the heat capacity in the vicinity of the critical temperature provides important information about the thermodynamic characteristics of the condensation process. In addition, the virial expansion offers a useful method for describing the equation of state and related thermodynamic quantities in terms of the particle density and thermal de Broglie wavelength.

The relevance of relativistic effects is not restricted to laboratory systems. Bose-Einstein condensate stars and other bosonic astrophysical objects have been considered in the literature at finite temperatures \cite{BE1,BE2,BE3,BE4}. In such systems, the particle mass and temperature may lead to relativistic contributions that modify the thermodynamic properties predicted by the non-relativistic theory. Consequently, a systematic analysis of relativistic corrections is necessary to assess their influence on the equation of state and the thermal behavior of these systems.

The purpose of the present work is to investigate the first relativistic corrections to the thermodynamic properties of a Bose-Einstein gas within the framework of relativistic kinetic theory. We begin by expanding the equilibrium distribution function in powers of the inverse speed of light and deriving the corresponding expressions for the particle number density, pressure, and specific internal energy. We then analyze the heat capacity at constant volume below, at, and above the Bose-Einstein condensation temperature, emphasizing the continuity of the specific heat in the non-ultrarelativistic regime. Finally, we develop a virial expansion to obtain the relativistic corrections to the equation of state and to several thermodynamic quantities, including the specific internal energy, specific heat, chemical potential, enthalpy, and entropy. These results provide a systematic description of relativistic effects in Bose-Einstein gases and may contribute to the understanding of finite-temperature bosonic astrophysical systems.

\section{Relativistic corrections to quantum distributions}\lb{s.2}

Let us consider a gas with particles of rest mass $m$ in the phase space described by the space-time coordinates $(x^\mu)=(x^0=ct,\bx)$ and the momentum four-vector $(p^\mu)=(p^0,\bp)$. The momentum four-vector length is $mc$, so that the constraint $p^0=\sqrt{\vert\bp\vert+m^2c^2}$ holds. 

The state of the gas in the phase space is characterized by the one-particle distribution function $f(x^\mu,p^\mu)=f(\bx,\bp,t)$, such that $f(\bx,\bp,t)d^3xd^3v$ gives, at time $t$ the number of particles in the volume element $d^3x$ around $\bx$ and with momenta in the range $d^3p$ around $\bp$.

The space-time evolution of the one-particle distribution function $f(\bx,\bp,t)$ in the phase space is determined by the Boltzmann equation (see e.g. \cite{CK}). The equilibrium distribution function follows from the consideration that the net number of particles that enter and leave the volume element in the phase space  vanishes. For a quantum  gas, the equilibrium distribution function was determined by J\"uttner in 1928 \cite{JT} and reads (here we write the expression  given in   the book \cite{CK}) 
\ben\lb{qt1}
f(\bx,\bp,t)=\frac{1}{h^3}\frac1{\exp\left(\frac{g_{\mu\nu}p^\mu U^\mu-\mu}{kT}\right)-\epsilon}.
\een
In the above equation, $h$ is Planck's constant, $T$ the absolute temperature, $k$ the Boltzmann constant and $\epsilon$ a parameter, that assumes the values $\epsilon=0$ for Maxwell-Boltzmann statistics, $\epsilon=+1$ for Bose-Einstein statistics, and $\epsilon=-1$ for Fermi-Dirac statistics. Furthermore, $g_{\mu\nu}$  denotes the metric tensor, $U^\mu$ the hydrodynamic four-velocity and $\mu$ the chemical potential per particle (Gibbs free energy).

 The components of the particle four-velocity  are given by \cite{Wein}
\ben\lb{qt2a}
(u^\mu)=\left(\frac{dx^0}{d\tau}, \frac{dx^i}{d\tau}\right)=\left(c\frac{dt}{d\tau},v_i\frac{dt}{d\tau}\right),
\een
and  obtained from  the expression that follows from the line element   
\ben\lb{qt2b}
\left(\frac{d\tau}{dt}\right)^2=g_{00}+\frac2cg_{0i}v^i+\frac1{c^2}g_{ij}v^iv^j.
\een

In Minkowski space-time, the components of the metric tensor are: $g_{00}=1$, $g_{0i}=0$ and $g_{ij}=-\delta_{ij}$, thus, up to the $1/c^4$ order in the speed of light, we have 
\ben\lb{qt3}
u^0=c\left[1+\frac{v^2}{2c^2}+\frac{3v^4}{8c^4}\right],
\qquad
u^i=\frac{v_iu^0}c.
\een

In a local Lorentz frame, where the gas is seen as an isotropic body for an observer that moves with the velocity of the gas, the components of the hydrodynamic four-velocity reduce to $(U^\mu)=(c,{\bf0})$. In this case, by taking  $p^\mu=m u^\mu$ we can expand the equilibrium distribution function (\ref{qt1}) up to $1/c^4$ order terms, yielding
\ben\lb{qt4}
f=\frac{1}{h^3}\frac1{e^{\frac{mv^2}{2kT}-\frac{\mu_*}{kT}}-\epsilon}
\left\{ 1-\frac{m}{kT}\frac{e^{\frac{mv^2}{2kT}-\frac{\mu_*}{kT}}}{{e^{\frac{mv^2}{2kT}-\frac{\mu_*}{kT}}-\epsilon}}\left[\left(\frac{3  v^4}{8 c^2}+\frac{5  v^6}{16 c^4}\right)-\frac{m}{kT}\frac{9  v^8}{128 c^4}\frac{e^{\frac{mv^2}{2kT}-\frac{\mu_*}{kT}}+\epsilon}{e^{\frac{mv^2}{2kT}-\frac{\mu_*}{kT}}-\epsilon}\right]\right\}.
\een
Here we have introduced the abbreviation $\mu_\star=\mu-mc^2$, for the difference of the chemical potential per particle $\mu$ and the particle rest energy $mc^2$. 

\section{Fields in equilibrium}\lb{s.3}

We shall determine the fields in equilibrium corresponding to the particle number density $n$, internal energy per particle $e$ and pressure $p$ from the components of the particle four-flow $N^\mu$ and of the energy-momentum tensor $T^{\mu\nu}$, which are defined in terms of the one-particle distribution function by (see e.g. \cite{CK})
\ben\lb{qt5}
N^\mu=m^3c\int u^\mu  f\frac{d^3 u}{u_0},
\qquad T^{\mu\nu}=m^4c\int u^\mu u^\nu f\frac{d^3 u}{u_0}.
\een 
In a local Lorentz rest frame we have $N^0=cn$, $T^{00}=ne$ and $T^{ij}=p\delta_{ij}$.

The transformation $d^3u\rightarrow d^3v$ up to the $1/c^4$ order is given by (see \cite{GK1})
\ben\lb{qt6}
\frac{ d^3 u}{u_0}=
\left[1+\frac{5v^2}{2c^2}+\frac{35v^4}{8c^4}\right]\frac{d^3v}{u^0}=
\left[1+\frac{2v^2}{c^2}+\frac{3v^4}{c^4}\right]\frac{d^3v}{c}.
\een

The  integrations proceed as follows:   we insert  the   distribution function (\ref{qt4}) and the integration element (\ref{qt6}) into (\ref{qt5}),  introduce  the  spherical coordinates $(v,\theta,\varphi)$, and  express the integral element as $d^3v=v^2\sin\theta d\theta d\varphi dW$, where $0\leq v<\infty$, $0\leq\theta\leq\pi$ and $0\leq\varphi\leq2\pi$.  The results of  the integrations are given below where we have introduced the integral 
\ben\lb{qt7a}
I_n\left(\frac{\mu_\star}{kT}\right)=\frac{1}{\Gamma(n)}\int_0^\infty\frac{x^{n-1}}{e^{x-\frac{\mu_\star}{kT}}-\epsilon}dx.
\een

From the integration of the time component of the particle four-flow, we get the expression for the particle number density
\ben\lb{qt8b}
n=\frac{N^0}c=\left(\frac{2\pi mkT}{h^2}\right)^\frac32I_{\frac32}(\frac{\mu_\star}{kT})\left[1+\frac{15kT}{8mc^2}\frac{I_{\frac52}(\frac{\mu_\star}{kT})}{I_{\frac32}(\frac{\mu_\star}{kT})}+\frac{105k^2T^2}{128m^2c^4}\frac{I_{\frac72}(\frac{\mu_\star}{kT})}{I_{\frac32}(\frac{\mu_\star}{kT})}\right].
\een

Similarly, the integration of the time and spatial components of the energy-momentum tensor $T^{00}$ and $T^{ij}$, leads to
\ben\lb{qt9a}
&&T^{00}=ne=mc^2\left(\frac{2\pi mkT}{h^2}\right)^\frac32I_{\frac32}(\frac{\mu_\star}{kT})\left[1+\frac{27kT}{8mc^2}\frac{I_{\frac52}(\frac{\mu_\star}{kT})}{I_{\frac32}(\frac{\mu_\star}{kT})}+\frac{705k^2T^2}{128m^2c^4}\frac{I_{\frac72}(\frac{\mu_\star}{kT})}{I_{\frac32}(\frac{\mu_\star}{kT})}\right],
\\\lb{qt9b}
&&T^{ij}=p\delta_{ij}=\left(\frac{2\pi mkT}{h^2}\right)^\frac32kTI_{\frac52}(\frac{\mu_\star}{kT})\left[1+\frac{15kT}{8mc^2}\frac{I_{\frac72}(\frac{\mu_\star}{kT})}{I_{\frac52}(\frac{\mu_\star}{kT})}+\frac{105k^2T^2}{128m^2c^4}\frac{I_{\frac92}(\frac{\mu_\star}{kT})}{I_{\frac52}(\frac{\mu_\star}{kT})}\right]\delta_{ij}.
\een

From now on, we shall write only terms up to the first relativistic correction related with $kT/mc^2$. By considering the expression for  the particle number density (\ref{qt8b}), we can write the internal energy $ne_\star=n(e-mc^2)$ and the pressure $p$ from (\ref{qt9a}) and (\ref{qt9b}) up to $kT/mc^2$ terms , respectively as
\ben\lb{qt10a}
&&ne_\star=\frac{3kT}{2}\left(\frac{2\pi mkT}{h^2}\right)^\frac32I_{\frac52}(\frac{\mu_\star}{kT})\left[1+\frac{25kT}{8mc^2}\frac{I_{\frac72}(\frac{\mu_\star}{kT})}{I_{\frac52}(\frac{\mu_\star}{kT})}\right],
\\\lb{qt10b}
&&p=nkT\frac{I_{\frac52}(\frac{\mu_\star}{kT})}{I_{\frac32}(\frac{\mu_\star}{kT})}\left[1+\frac{15kT}{8mc^2}\left(\frac{I_{\frac72}(\frac{\mu_\star}{kT})}{I_{\frac52}(\frac{\mu_\star}{kT})}-\frac{I_{\frac52}(\frac{\mu_\star}{kT})}{I_{\frac32}(\frac{\mu_\star}{kT})}\right)\right].
\een

The specific internal energy follows from (\ref{qt10a}) and (\ref{qt8b}), yielding
\ben\lb{qt10c}
e_\star=\frac{3}{2}kT\frac{I_{\frac52}(\frac{\mu_\star}{kT})}{I_{\frac32}(\frac{\mu_\star}{kT})}\left[1+\frac{5kT}{4mc^2}\left(\frac52\frac{I_{\frac72}(\frac{\mu_\star}{kT})}{I_{\frac52}(\frac{\mu_\star}{kT})}-\frac{I_{\frac52}(\frac{\mu_\star}{kT})}{I_{\frac32}(\frac{\mu_\star}{kT})}\right)\right].
\een

From the above equations, one may obtain the first relativistic corrections to the fields of particle number density, pressure and specific internal energy corresponding to the Maxwell-Boltzmann, Fermi-Dirac and Bose-Einstein statistics, by considering in the integral (\ref{qt7a}) the parameter $\epsilon$ equal to 0, -1 or +1, respectively. For the Maxwell-Boltzmann and Fermi-Dirac statistics, these corrections are well-known in the literature (see e.g. \cite{CK}). Here, we are interested in analyzing the relativistic corrections to these fields in the Bose-Einstein statistics, and in this case,  the integral (\ref{qt7a}) becomes  the Bose-Einstein  functions \cite{Patr}   
\ben\lb{qt7b}
g_n(z)=\frac1{\Gamma(n)}\int_0^\infty\frac{x^{n-1}}{z^{-1}e^x-1}dx=z+\frac{z^2}{2^n}+\frac{z^3}{3^n}+\frac{z^4}{4^n}+\dots,\qquad \hbox{where}\qquad z=\exp\left(\frac{\mu_*}{kT}\right).
\een
For $z=1$, we have  $g_n(1)=\zeta(n)$ where $\zeta(n)$ is the  Riemann zeta function.

\section{Heat capacity at Bose-Einstein condensation }

In this section, we shall analyze the heat capacity of a Bose-Einstein gas near the critical temperature $T_c$, where  Bose-Einstein condensation occurs.

For temperatures $T$ below  the critical temperature $0\leq T\leq T_c$ we have $z\simeq1$, i.e., the chemical potential is practically zero  and  $g_n(1)\equiv\zeta(n)$. The heat capacity at constant volume follows from (\ref{qt10a}) and reads
\ben\lb{cv1}
C_v=\left(\frac{\partial ne_\star}{\partial T}\right)_v=\frac{15}4 k\left(\frac{2\pi mkT}{h^2}\right)^\frac32\zeta(5/2)\left\{1+\frac{35kT}{8mc^2}\frac{\zeta(7/2)}{\zeta(5/2)}\right\},
\een

For temperatures greater than the critical temperature $T\geq T_c$, we take the temperature derivative of the particle number density (\ref{qt8b}) at constant volume and obtain
\ben
\frac{\partial z}{\partial T}=-\frac{3z}{2T}\frac{g_{\frac32}(z)}{g_{\frac12}(z)}\left\{1+\frac{15kT}{8mc^2}\left[\frac53\frac{g_{\frac52}(z)}{g_{\frac32}(z)}-\frac{g_{\frac32}(z)}{g_{\frac12}(z)}\right]\right\},
\een
which is used to calculate the heat capacity at constant volume for $T\geq T_c$:
\ben\lb{cv2}
C_v=\left(\frac{\partial ne_\star}{\partial T}\right)_v=\frac{15}{4}k\left(\frac{2\pi mkT}{h^2}\right)^\frac32g_{\frac32}(z)\left\{\frac{g_{\frac52}(z)}{g_{\frac32}(z)}-\frac35\frac{g_{\frac32}(z)}{g_{\frac12}(z)}+\frac{9kT}{8mc^2}\left[\frac{g_{\frac72}(z)}{g_{\frac32}(z)}-\frac67\frac{g_{\frac52}(z)}{g_{\frac12}(z)}+\frac9{35}\frac{g_{\frac32}(z)^2}{g_{\frac12}(z)^2}\right]\right\}.
\een

The specific heat capacities follow from (\ref{qt8b}), (\ref{cv1}) and (\ref{cv2}), yielding
\ben\lb{cv3a}
c_v=\frac{C_v}{n}=\frac{15 k }{4}\left[\frac{\zeta(5/2)}{\zeta(3/2)}+\frac{15kT}{8mc^2}\left(\frac{7\zeta(7/2)}{3\zeta(3/2)}-\frac{\zeta(5/2)^2}{\zeta(3/2)^2}\right)\right],\qquad T\leq T_c,
\\\lb{cv3b}
c_v=\frac{C_v}{n}=\frac{15}{4}k\left\{\frac{g_{\frac52}(z)}{g_{\frac32}(z)}-\frac35\frac{g_{\frac32}(z)}{g_{\frac12}(z)}+\frac{15kT}{8mc^2}\left[\frac73\frac{g_{\frac72}(z)}{g_{\frac32}(z)}-\frac{g_{\frac52}(z)^2}{g_{\frac32}(z)^2}+\frac{3}{5}\frac{g_{\frac32}(z)^2}{g_{\frac12}(z)^2}-\frac75\frac{g_{\frac52}(z)}{g_{\frac12}(z)}\right]\right\},\qquad T\geq T_c.
\een

At the critical temperature $T_c$, the limiting values of the specific heat capacities agree and are given by
\ben
c_v(T_c^-)=c_v(T_c^+)=\left(1.92567  + \underline{5.22201 \frac{kT_c}{mc^2}}\right)k,
\een
since at $T=T_c$, the chemical potential $\mu_*\approx0$ and  $g_{n}(z)\rightarrow\zeta(n)$. The  underlined term refers to the first relativistic correction to the well-known non-relativistic value. Here we note that the specific heat capacity is continuous at the critical temperature, which is not the case in   ultrarelativistic regime where  a jump  of order ${c_v}(T_c^-)-{c_v}(T_c^+)\approx 6.5768k$ occurs. This was first reported by Landsberg and Dunning-Davis \cite{LDD} (see also \cite{CK}). 

For $T \gg T_c$,  the specific heat decreases toward the classical value $3k/2$ plus a relativistic correction of the Maxwell-Boltzmann statistics (see e.g. \cite{CK}), namely
\ben
c_v=\frac32\left(1+ \frac{5k T}{2mc^2 }\right)k.
\een 

In general, the relativistic corrections near the non-relativistic regime are very small, but there are some papers in the literature that consider Bose-Einstein condensate stars at finite temperature (see e.g. \cite{BE1,BE2,BE3,BE4}). In this case we can take the mass of the condensate particle as twice the nucleon mass $m=2m_n=3.34986 \times10^{-27}$kg --
by considering  the particles of Bose-Einstein condensate stars as equivalent of a Cooper pair -- and the temperature in the range of 1 to 4 $\times 10^{11}$K. With these values, the specific heat at the critical temperature becomes $c_v=2.02128 k$, which is about 5\% greater than the one without the relativistic correction $c_v=1.92567k$. For $T \gg T_c$ the specific heat is about 4.5\% greater than the well-known non-relativistic result $c_v=3k/2$.

\section{The virial equation of state}

 In order to analyze the virial equation of state, we introduce a parameter $\lambda=h/\sqrt{2\pi mkT}$, which is associated to de Broglie wavelength, and express  the particle number density (\ref{qt8b}) in terms of the specific volume $v=1/n$. For that end we consider the expansion in the integral (\ref{qt7b}) and write
\ben\lb{v1}
\frac{\lambda^3}{v}=z+\frac{z^2}{2\sqrt2}+\frac{z^3}{3\sqrt3}+\underline{\frac{15kT}{8mc^2}\left[z+\frac{z^2}{4\sqrt2}+\frac{z^3}{9\sqrt3}\right]}.
\een

From (\ref{v1}) we can determine $z$ as a function of $\lambda^3/v$ through the inversion  of the series  up to $(\lambda^3/v)^3$ and up to the first relativistic correction $kT/mc^2$ and get
\ben\lb{v2}
z=\left(\frac{\lambda^3}{v}\right)-\frac1{2\sqrt2}\left(\frac{\lambda^3}{v}\right)^2+\left(\frac14-\frac1{2\sqrt3}\right)\left(\frac{\lambda^3}{v}\right)^3-\underline{\frac{15kT}{8mc^2}\left[\left(\frac{\lambda^3}{v}\right)-\frac5{4\sqrt2}\left(\frac{\lambda^3}{v}\right)^2+\left(1-\frac{11}{9\sqrt3}\right)\left(\frac{\lambda^3}{v}\right)^3\right]}.
\een

The virial equation of state follows from (\ref{qt10b}) by taking into account the virial expansion (\ref{v2}), which, by considering terms up to  $(\lambda^3/v)^2$ reads
\ben\lb{v3}
\frac{pv}{kT}=1-\frac1{4\sqrt2}\left(\frac{\lambda^3}{v}\right)+\left(\frac18-\frac2{9\sqrt3}\right)\left(\frac{\lambda^3}{v}\right)^2+\underline{\frac{kT}{mc^2}\left[\frac{45}{64\sqrt2}\left(\frac{\lambda^3}{v}\right)+\left(\frac{10}{9\sqrt3}-\frac{45}{64}\right)\left(\frac{\lambda^3}{v}\right)^2\right]}.
\een
 The first relativistic correction to the virial equation of state of a Bose-Einstein gas is given by the underlined term in (\ref{v3}).

The dependence of the specific internal energy on the virial coefficients results from  (\ref{qt10c}) and the virial expansion (\ref{v2}), yielding
\ben\lb{v4}
e_\star=\frac32kT\left\{1-\frac1{4\sqrt2}\left(\frac{\lambda^3}{v}\right)+\left(\frac18-\frac2{9\sqrt3}\right)\left(\frac{\lambda^3}{v}\right)^2+\underline{\frac{5kT}{6mc^2}\left[1-\frac{3}{8\sqrt2}\left(\frac{\lambda^3}{v}\right)+\left(\frac{25}{27\sqrt3}-\frac{75}{128}\right)\left(\frac{\lambda^3}{v}\right)^2\right]}\right\},
\een
from which  the virial expansion of the specific heat follows
\ben\no
&&c_v=\left(\frac{\partial e_\star}{\partial T}\right)_v=\frac32k\left\{1+\frac1{8\sqrt2}\left(\frac{\lambda^3}{v}\right)+\left(\frac4{9\sqrt3}-\frac14\right)\left(\frac{\lambda^3}{v}\right)^2\right.
\\\lb{v5}
&&\qquad\left.\underline{+\frac{5kT}{3mc^2}\left[1+\frac{3}{32\sqrt2}\left(\frac{\lambda^3}{v}\right)+\left(\frac{75}{256}-\frac{25}{54\sqrt3}\right)\left(\frac{\lambda^3}{v}\right)^2\right]}\right\}.
\een

In the above equations, the underlined terms refer to the first relativistic corrections to the corresponding non-relativistic quantities, which can be found in the literature (see e.g  \cite{Patr}).

Here, we can also determine the virial expansion and the corresponding relativistic correction of some other thermodynamic quantities. We begin by determining  the chemical potential, which  can be obtained from (\ref{v2}) by considering the series of $\mu_\star=kT\ln z$ up to $(\lambda^3/v)^2$ and $kT/mc^2$:
\ben\lb{v6}
\mu_*=kT\left\{\ln\left(\frac{\lambda^3}{v}\right)-\frac1{2\sqrt2}\left(\frac{\lambda^3}{v}\right)+\left(\frac{3}{16}-\frac{\sqrt3}{9}\right)\left(\frac{\lambda^3}{v}\right)^2-\frac{15kT}{8mc^2}\left[1-\frac{3}{4\sqrt2}+\left(\frac9{16}-\frac8{9\sqrt3}\right)\left(\frac{\lambda^3}{v}\right)^2\right]\right\}.
\een

From the above equations, one obtains the virial expressions and the corresponding relativistic corrections for the thermodynamic potentials, enthalpy per particle $h$, and entropy per particle $s$, which follow from (\ref{v3}), (\ref{v4}),  (\ref{v6}) and are given by
\ben\no
&&h=e_\star+pv=\frac52kT\left\{1-\frac1{4\sqrt2}\left(\frac{\lambda^3}{v}\right)+\left(\frac18-\frac2{9\sqrt3}\right)\left(\frac{\lambda^3}{v}\right)^2\right.
\\
&&\qquad\left.+\frac{kT}{2mc^2}\left[1+\frac{15}{16\sqrt2}\left(\frac{\lambda^3}{v}\right)+\left(\frac{49}{27\sqrt3}-\frac{147}{128}\right)\left(\frac{\lambda^3}{v}\right)^2\right]\right\},
\\\no
&&s=\frac1T\left(e_\star+pv-\mu_\star\right)=k\left\{\frac52-\ln\left(\frac{\lambda^3}{v}\right)-\frac1{8\sqrt2}\left(\frac{\lambda^3}{v}\right)+\left(\frac18-\frac2{9\sqrt3}\right)\left(\frac{\lambda^3}{v}\right)^2\right.
\\
&&\left.\qquad+\frac{25kT}{8mc^2}\left[1-\frac3{40\sqrt2}\left(\frac{\lambda^3}{v}\right)+\left(\frac{26}{135\sqrt3}-\frac{39}{320}\right)\left(\frac{\lambda^3}{v}\right)^2\right]\right\}.
\een

\section{CONCLUSIONS}

In this work, we have investigated the relativistic corrections to the thermodynamic properties of a Bose-Einstein gas within the framework of relativistic kinetic theory. By expanding the equilibrium distribution function in powers of the inverse speed of light, we have obtained expressions for the particle number density, internal energy, and pressure, including the first relativistic corrections of order $(kT/mc^2)$.

The heat capacity at constant volume has been analyzed in the vicinity of the Bose-Einstein condensation temperature. We have shown that the specific heat is continuous at the critical temperature in the non-ultrarelativistic regime, and that its limiting value includes a relativistic contribution proportional to $(kT_c/mc^2)$. In the high-temperature limit, the specific heat approaches the classical Maxwell-Boltzmann value with its corresponding relativistic correction.

The numerical estimates for Bose-Einstein condensate stars indicate that relativistic effects may produce appreciable modifications to the thermodynamic properties. In particular, for the considered particle mass and temperature range, the specific heat at the critical temperature is approximately 5\% larger than its non-relativistic value, while in the high-temperature limit the increase is about 4.5\%.

Furthermore, we have developed a virial expansion of the equation of state and derived the corresponding relativistic corrections to the specific internal energy, specific heat, chemical potential, enthalpy, and entropy. These expressions provide a systematic extension of the non-relativistic results and describe the influence of relativistic effects on the thermodynamic behavior of a Bose-Einstein gas at finite density.

The results obtained in this work demonstrate that, although relativistic corrections are generally small in the non-relativistic regime, they can become relevant in the description of Bose-Einstein condensate stars at finite temperature. The present analysis therefore provides a useful theoretical framework for investigating the thermodynamic properties of relativistic Bose-Einstein gases and their possible astrophysical applications.

\end{document}